\documentclass{paper}
\usepackage{amsfonts,amsmath,latexsym,amscd,graphicx,amssymb}

\newcommand{\be}{\begin{equation}}
\newcommand{\ee}{\end{equation}}
\newcommand{\ba}{\begin{eqnarray}}
\newcommand{\ea}{\end{eqnarray}}

\newcommand{\pa}{\partial}
\newcommand{\f}{\frac}

\title {Elliptic and hyperelliptic functions describing the particle
motion beneath small-amplitude water waves with constant
vorticity}

\author{\normalsize Delia IONESCU-KRUSE\\
\normalsize Institute of Mathematics of
the Romanian Academy,\\
\normalsize P.O. Box 1-764, RO-014700, Bucharest,
 Romania\\
\normalsize E-mail: Delia.Ionescu@imar.ro\\[10pt]}

 \date{}

\begin{document}
\maketitle

\begin{abstract}
We provide analytic solutions of the nonlinear differential
equation system describing the particle paths below
small-amplitude periodic gravity waves travelling  on  a constant
vorticity current. We show that  these paths are not closed
curves. Some solutions can be expressed in terms of Jacobi
elliptic functions, others in terms of hyperelliptic functions. We
obtain new kinds of particle paths. We make some remarks on the
stagnation points which could appear in the fluid due to the
vorticity.
\end{abstract}

\section{Introduction}

The present work is confined to two-dimensional water waves with
constant vorticity. To gain insight into the motion in such waves
we determine analytically the trajectories beneath small-amplitude
water waves in  constant vorticity flows. In order to get this
Lagrangian feature of the flow, that is, the evolution of
individual water particles, we will firstly find a solution of the
Eulerian system of equations within the framework of small
amplitude waves and then we will calculate the  solutions of the
nonlinear differential equations system which describes the
particle motion.

Many of the theoretical results concerning  waves on water make
the initial assumption of irrotational flow. There are
circumstances in which this is well justified but there are cases
where it is inappropriate. Waves with vorticity are commonly seen
in nature, for example, in shear currents. Tidal flow is a
well-known example when constant vorticity flow is an appropriate
model (see Da Silva T. A. and Peregrine D. H. \cite{dasilva}). For
a  discussion of the  physical relevance of flows with constant
vorticity see also  \cite{c2011}. In 1802, Gerstner
\cite{gerstner} constructed an explicit example of a periodic
travelling wave in water of infinite depth with a specific
non-constant vorticity\footnote{This solution was independently
re-discovered later by Rankine \cite{rankina}. Modern detailed
descriptions of this wave are given in the recent papers
\cite{c2001a} and \cite{henry4}.}. The fact that this flow is very
special is confirmed also by the fact that this is the only steady
flow satisfying the constraint of constant pressure along the
streamlines cf. \cite{kalisch}. Gerstner's wave is a
two-dimensional wave which adopts the Lagrangian viewpoint,
describing the evolution of individual water particles. Its
surface profile is symmetric \cite{c2001a}. Beneath Gerstner's
wave it is possible to have a motion of the fluid where all
particles describe circles with a depth-dependent radius
\cite{c2001a}, \cite{henry4}.

 In 1934, Dubreil-Jacotin
\cite{dubreil} considered the problem of the existance of steady
periodic water waves with general vorticity. She proved the
existence of large classes of small-amplitude water waves with
vorticity. For large-amplitude water waves with vorticity, in
2004, Constantin and Strauss \cite{CS} proved that, for an
arbitrary vorticity distribution and for a given $c>0$ and
relative mass flux $p_0$, there is a global continuum of steady
periodic waves travelling at speed $c$ in water of finite depth
and such that the horizontal component of the velocity $u<c$
throughout the fluid. The continuum contains waves with $u$
arbitrarily close to the wave speed $c$. The existence of global
continua of smooth solutions for the related problem of periodic
waves of infinite depth was proved by Hur \cite{hur}.   The
vorticity does not destroy the symmetry. The construction in
\cite{CS} assumes that the wave profiles are symmetric. Constantin
and Escher \cite{ce1}, \cite{ce2} proved that the symmetry of the
wave profile is not a hypothesis but rather a conclusion when the
wave profile is monotone between crest and trough and the
vorticity is positive and non-increasing with greater depth. Free
from restrictions on the vorticity but requiring a quite precise
knowledge of all streamlines in the fluid, Hur proved in
\cite{hur2} that if the wave profile is monotone near the trough
and every streamline has a single minimum per wavelength located
below the trough, then the steady periodic water waves of finite
depth are symmetric. Constantin, Ehrnst\" om and Wahl\' en
\cite{cew} showed  that for an arbitrary vorticity distribution, a
steady periodic water wave with a profile that is monotone between
crests and troughs has to be symmetric.

 Another  remarkable feature of  the rotational steady
 waves is that they could
 contain stagnation points. If $(u,v)$ denotes
the velocity field and $c$ the constant speed  of the wave, then a
point where $u=c$ and $v=0$ is called stagnation point. There are
interesting problems related to the so-called extreme waves: these
are waves with the stagnation points at their crests. Varvaruca
\cite{varvaruca1} proved for a certain class of vorticity
functions, the existence of extreme waves and showed that at such
a stagnation point the profile of the wave has either a corner of
$120^0$ or a horizontal tangent. For a recent survey of different
aspects of the theory of steady water waves with vorticity see
\cite{strauss}.

This paper has interest in finding information about the flow
below  water waves  with constant vorticity, more precisely, we
will investigate how the presence of vorticity influences the
particle paths. Throughout the hydrodynamics literature, it has
been quite common to assume that beneath an irrotational periodic
two-dimensional travelling water wave, the particles trace closed,
circular or elliptic, orbits. (see for example \cite{lamb},
\cite{lighthill}, \cite{debnath}, \cite{johnson-carte}). While in
this first approximation  all particle paths appear to be closed,
Constantin and Villari showed in \cite{cv}, using phase-plane
considerations for the nonlinear system describing the particle
motion, that in linear irrotational periodic gravity water waves
no particles trajectory is actually closed, unless the free
surface is flat. Similar results hold for the particle
trajectories in irrotational deep-water (see Constantin,
Ehrnstr\"{o}m and Villari \cite{cev}), and in irrotational shallow
water (see Ionescu-Kruse  \cite{io} and \cite{io2}, Section 5.1).
Ionescu-Kruse \cite{io}, \cite{io2} obtained the exact solutions
of the nonlinear differential equation system which describes the
particle motion in small-amplitude shallow water waves  and showed
that there does not exist a single pattern for all particles:
depending on the strength of the underling uniform current, some
particle trajectories are undulating curves to the right, or to
the left, others are loops with forward drift, and others are not
physically acceptable, in the last case it seems necessary to
study the
full nonlinear problem.\\
 For the full nonlinear  problem, Constantin proved in
 \cite{c2007}, by analyzing a free
boundary problem for harmonic functions in a planar domain, that
all water  particles in Stokes waves display a forward
 drift. For an extension of the
investigation in \cite{c2007} to deep-water Stokes waves see Henry
\cite{henry}.  In a very recent paper  \cite{cs2010},  Constantin
and Strauss  recovered the results in \cite{c2007}  by a simpler
approach and they  also investigated the effect of an underlying
current on the paths of the particles.  While in periodic waves
within a period each particle experiences a backward-forward
motion with a forward drift, Constantin and Escher
 showed in \cite{CE2} that in a solitary water wave there is no backward motion:
 all particles move in the direction of wave propagation
 at a positive speed, the direction being upwards
 or downwards if the particle precedes, respectively, does not
 precede the wave crest.

There have also been some studies of particle paths for rotational
waves. Within  the linear theory, by using phase-plane
considerations for the nonlinear system describing the particle
motion,  Ehrnstr\" om and Villari \cite{ev}  found that for
positive constant  vorticity, the behavior of the streamlines is
the same as for the irrotational waves, though the physical
particle paths behave differently if the size of the vorticity is
large enough. For negative vorticity they showed that
 in a frame moving with the wave, the fluid
contains a cat's-eye vortex (see \cite{majda}, Ex. 2.4). The paper
\cite{w2} by Wahl\' en which contains an existence result for
small-amplitude solutions, based on local bifurcation theory,
showed also that the predictions for negative vorticity \cite{ev}
in the linear theory are true. We mention that an alternative
approach to the existence result in \cite{w2} for small-amplitude
steady waves with constant vorticity was very recently proposed by
Constantin and Varvaruca \cite{cv2}.
 Beside the
phase-plane analysis, the exact solutions of the nonlinear system
describing the particle motion, allow a better understanding of
the dynamics. For small-amplitude shallow-water waves with
vorticity and background flow Ionescu-Kruse found in \cite{io2}
the exact solutions and showed that depending on the relation
between the initial data and the constant vorticity some particles
trajectories are undulating curves to the right, or to the left,
others are loops with forward drift, or with backward drift,
others can follow peculiar shapes (see \cite{io2}, Fig. 7e).

Removing the shallow-water restriction, in the present paper we
provide explicit solutions for the  nonlinear system describing
the motion of the particles beneath small-amplitude gravity waves
which propagate on
the surface of a constant vorticity flow. \\
In Section 2 we recall the governing equations for gravity water
waves.\\
 In Section 3 we present their nondimensionalisation and
scaling. We present two different scalings, in one the constant
vorticity $\omega_0$  is scaled whereas in another one $\omega_0$
remains unscaled. We choose $x$ and $z$ the space coordinates,
thus, the sign of the constant vorticity $\omega_0$ is opposite to
the sign of the constant
vorticity considered if $x$ and $y$ are chosen the space coordinates. \\
In Section 4 we obtain the periodic travelling solutions of the
considered linearized problems  (see (\ref{solrotconst}),
respectively (\ref{solrotconst2})), and the speed of propagation
of the linear wave  $c$  (see (\ref{c}), respectively
(\ref{c''})). The solutions are also written in the original
physical variables: see
 (\ref{c1}), (\ref{solrotconst'}),  respectively (\ref{cphys}),
(\ref{solrotconst2'}). We observe that the speed of the wave and
the pressure have different expressions in the two
linearizations. \\
In Section 5 we find the solutions of the nonlinear differential
equation systems (\ref{diff2}) and we describe the possible
particle trajectories beneath constant vorticity water waves.  In
the study of the system (\ref{diff2}) it is interesting to observe
that, for the first  linearization, that is, the one made around
still water in which the constant vorticity $\omega_0$ is scaled,
the sign of the wave speed $c$ will influence the sign of the
parameter $A$ which appears in the components $u$ and $v$ of the
velocity field. Thus, if  we consider left-going waves $c<0$, we
get $A<0$ and if
 we
consider right-going waves $c>0$, we obtain $A>0$. For the second
linearization, that is, the one made  around a laminar flow
characterized by $u=\omega_0 z+\alpha,\, v=0$, $\alpha$ being a
constant, we obtain that, independent of the sign of $\omega_0$,
the sign of $A$ depends on the sign of
$c-h_0\omega_0-\sqrt{gh_0}\alpha$, where $h_0$ is the finite depth
and $g$ the constant gravitational acceleration.  The expression
$c-h_0\omega_0-\sqrt{gh_0}\alpha$ could be regarded as "the speed"
of a wave which can be left-going or right-going.
In the study
 of the system (\ref{diff2})
 a peakon-like trajectory (\ref{sol0}) comes up (see also Ionescu-Kruse
 \cite{io5}). This solution
contains the arctanh$(\cdot)$ function, having a vertical
asymptote in the positive direction (Figure 2). For this solution
$u=c$ and a stagnation point in the fluid appear only for
$t\rightarrow \pm\infty$, where the path of the particle has a
horizontal tangent. The other solutions of the system
(\ref{diff2}) are given by (\ref{sol2}). We show  these solutions
are not closed curves. Some of these solutions can  be expressed
with the aid of the Jacobi elliptic functions, others are
expressed with the aid of the hyperelliptic functions.  We draw
some of the curves  obtained for different values of the
parameters (see Figure 3, Figure 4, Figure 5). At the end, we make
some remarks on the stagnation points inside the fluid.

\section{The water wave problem}
The two-dimensional gravity waves on constant vorticity water of
finite depth are described by the following boundary value
problem:

\begin{equation}
\begin{array}{c}
\begin{array}{ll}
u_t+uu_x+vu_z=- p_x\\ \,\, v_t+uv_x+vv_z=- p_z-g\\
\end{array}
\quad \quad \quad \quad \textrm{ (EEs) }\\
\\

 \qquad \qquad u_x+v_z=0  \qquad \qquad \qquad \qquad \textrm{ (MC)  }\\
\\

 \qquad \qquad u_z-v_x=\omega_0  \qquad \qquad \qquad \quad \textrm{ (VE)  }\\
\\

\begin{array}{ll}
  v=\eta_t+u\eta_x \, \, \textrm{ on }\,
z=h_0+\eta(x,t)\\
\qquad \quad v=0 \, \, \textrm { on } z=0
\end{array}
\quad \,\,\, \textrm{ (KBCs) }\\
\\

\qquad
 p=p_0 \,  \textrm{ on } z=h_0+\eta(x,t)
  \quad \quad \textrm{ (DBC)} \end{array} \label{e+bc}
\ee  where $(u(x,z,t), v(x,z,t))$ is the velocity field of the
water - no motion takes place in the $y$-direction, $p(x,z,t)$
denotes the pressure, $g$ is the constant gravitational
acceleration, $p_0$ being  the constant atmospheric pressure and
$\omega_0$ is the constant vorticity. The water moves in a domain
with a free upper surface at $z=h_0+\eta(x,t)$, for a constant
$h_0>0$, and a flat bottom at $z=0$. We set the constant water
density $\rho=1$. See in the Figure 1
 an example of a linear shear flow with
 constant  vorticity $\omega=\textrm{const}:=\omega_0>0$.
 \\

 \hspace{2cm}\scalebox{0.65}{\includegraphics{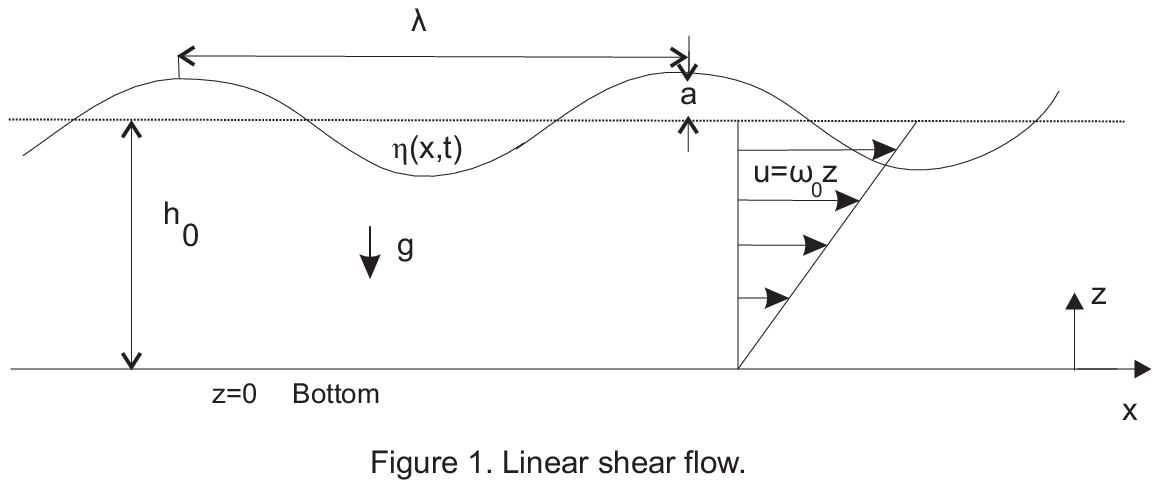}}
\\

\section {Non-dimensionalization and scaling}
We non-dimensionalize the set of equations (\ref{e+bc}) using the
undisturbed depth of the water $h_0$, as the vertical scale, a
typical wavelength $\lambda$, as the horizontal scale, and a
typical amplitude of the surface wave $a$ (for more details see
\cite{johnson-carte}). Thus, we define the set of non-dimensional
variables
\begin{equation}
\begin{array}{c}
x\mapsto\lambda x,  \quad z\mapsto h_0 z, \quad \eta\mapsto a\eta,
\quad t\mapsto\f\lambda{\sqrt{gh_0}}t,\\
 \\

  u\mapsto  \sqrt{gh_0}u,
\quad v\mapsto h_0\f{\sqrt{gh_0}}{\lambda}v,\\
 \\

 p\mapsto p_0+ g h_0(1-z)+ g h_0 p,
 \end{array} \label{nondim}\end{equation}
where, to avoid new notations, we have used the same symbols for
the non-dimensional variables  $x$, $z$, $\eta$, $t$, $u$, $v$,
$p$ on the right-hand side. The partial derivatives  $u_z$ and
$v_x$ will be then replaced by
\begin{equation}
u_z\mapsto\f {\sqrt{gh_0}}{h_0}u_z, \quad v_x\mapsto
h_0\f{\sqrt{gh_0}}{\lambda^2}v_x, \label{derivate}\end{equation}
and the natural scaling for the vorticity is \be \omega_0
\mapsto\f {\sqrt{gh_0}}{h_0}\omega_0, \label{nodim'}\ee where we
have used the same symbol for the non-dimensional $\omega_0$ on
the right-hand side.

 \noindent Therefore, in
non-dimensional variables (\ref{nondim}), (\ref{nodim'}), the
water-wave problem (\ref{e+bc}) becomes:
\begin{equation}
\begin{array}{cc}
u_t+uu_x+vu_z=- p_x&\\  \delta^2(v_t+uv_x+vv_z)=- p_z&\\
 u_x+v_z=0&\\
 u_z-\delta^2v_x=\omega_0&\\
v=\epsilon(\eta_t+u\eta_x) & \textrm{ on }\,
z=1+\epsilon\eta(x,t)\\
 p=\epsilon\eta&
\textrm{ on }\,
z=1+\epsilon\eta(x,t)\\
 v=0 &
\textrm { on } \, z=0
 \end{array}
\label{e+bc'} \end{equation}  where we have introduced the
amplitude parameter $\epsilon=\f a{h_0}$ and the shallowness
parameter $\delta=\f {h_0}{\lambda}$.

After the non-dimensionalization of  the system (\ref{e+bc}) let
us now proceed  with the scaling transformation. First we observe
that, on $z=1+\epsilon\eta$, both $v$ and $p$ are proportional to
$\epsilon$. This is consistent with the fact that as
$\epsilon\rightarrow 0$ we must have $v\rightarrow 0$ and
$p\rightarrow 0$. We can consider the following scaling of the
non-dimensional variables
\begin{equation} p\mapsto \epsilon p,\quad
u\mapsto\epsilon u,\quad v\mapsto\epsilon v
\label{scaling}\end{equation} where we avoided again the
introduction of a new notation. For this scaling of $u$ and $v$,
we also get \be
 \omega_0\mapsto
\epsilon \omega_0 \label{scaling'} \ee
 The water-wave
problem (\ref{e+bc}) writes in non-dimensional scaled variables
(\ref{nondim}), (\ref{nodim'}), (\ref{scaling}), (\ref{scaling'}),
as
\begin{equation}
\begin{array}{cc}
u_t+\epsilon(uu_x+vu_z)=- p_x&\\  \delta^2[v_t+\epsilon(uv_x+vv_z)]=- p_z&\\
 u_x+v_z=0&\\
  u_z-\delta^2v_x=\omega_0&\\
  v=\eta_t+\epsilon u\eta_x  \, & \textrm{ on }\,
z=1+\epsilon\eta(x,t)\\
p=\eta \, & \textrm{ on }\,
z=1+\epsilon\eta(x,t)\\
 v=0 \, &
\textrm { on } z=0
 \end{array}
\label{e+bc1''} \end{equation} By letting $\epsilon\rightarrow 0$,
$\delta$  being fixed, we obtain a linear approximation of
 the problem (\ref{e+bc1''}), that is,
\begin{equation}
\begin{array}{cc}
u_t+p_x=0&\\ \delta^2v_t+ p_z=0&\\
 u_x+v_z=0&\\
 u_z-\delta^2v_x=\omega_0&\\
v=\eta_t  \, & \textrm{ on }\,
z=1\\
  p=\eta \, & \textrm{ on }\,
z=1\\
 v=0 \, &
\textrm { on } z=0
\end{array}
\label{small} \end{equation} This linearization is used in
\cite{io}, \cite{io2} for irrotational and  constant vorticity
shallow water waves, in \cite{io3}, \cite{io4} for
capillary-gravity waves and in \cite{io5} for constant vorticity
gravity waves. \\
For constant vorticity flows all the streamlines are real-analytic
as proved recently in \cite{ce3}. Thus, for travelling water waves
with constant vorticity one can get the analytic validity of the
linearization (\ref{small}). In the case of irrotational water
waves, the rigorous analysis of the validity and relevance of  the
linearizations around some reference states is carried out in
\cite{lannes1}.

Instead of the scaling (\ref{scaling}), we can also consider the
following one
\begin{equation} p\mapsto \epsilon p,\quad
u\mapsto \omega_0z+\alpha+\epsilon u,\quad v\mapsto\epsilon v
\label{scaling2}\end{equation} with $\alpha$ constant. In this
case $\omega_0$ remains unscaled. Thus,  the water-wave problem
(\ref{e+bc}) writes in non-dimensional scaled variables
(\ref{nondim}), (\ref{nodim'}), (\ref{scaling2}), as
\begin{equation}
\begin{array}{cc}
u_t+\epsilon(uu_x+vu_z)+(\omega_0z+\alpha)u_x+\omega_0 v=- p_x&\\
 \delta^2[v_t+\epsilon(uv_x+vv_z)+(\omega_0 z+\alpha)v_x]=- p_z&\\
 u_x+v_z=0&\\
  u_z-\delta^2v_x=0&\\
  v=\eta_t+\epsilon u\eta_x +\epsilon \omega_0\eta\eta_x+(\omega_0 +\alpha)\eta_x \, & \textrm{ on }\,
z=1+\epsilon\eta(x,t)\\
p=\eta \, & \textrm{ on }\,
z=1+\epsilon\eta(x,t)\\
 v=0 \, &
\textrm { on } z=0
 \end{array}
\label{e+bc2} \end{equation} By letting $\epsilon\rightarrow 0$,
$\delta$  being fixed, we obtain a linear approximation of
 the problem (\ref{e+bc2}), that is,
\begin{equation}
\begin{array}{cc}
u_t+(\omega_0z+\alpha)u_x+\omega_0v+p_x=0&\\ \delta^2[v_t+(\omega_0 z+\alpha)v_x]+ p_z=0&\\
 u_x+v_z=0&\\
 u_z-\delta^2v_x=0&\\
v=\eta_t +(\omega_0 +\alpha)\eta_x \, & \textrm{ on }\,
z=1\\
  p=\eta \, & \textrm{ on }\,
z=1\\
 v=0 \, &
\textrm { on } z=0
\end{array}
\label{small2} \end{equation} We observe that the forth equation
in the system (\ref{small2}), which  represents the vorticity
equation, becomes in these scaled variables the vorticty equation
for an irrotational flow.\\
 The linearization (\ref{small2}) is
used in \cite{cv} for irrotational gravity waves, in \cite{cev}
for the corresponding deep-water waves, in \cite{henry2},
\cite{henry3} for capillary-gravity waves and in \cite{ev} for
gravity waves over finite depth with constant vorticity. This
linearization is around a laminar flow. Such shear flows are
characterized by the flat surface, $z=1$, corresponding to
$\eta=0$, $p=0$, $v=0$ and $u=\omega_0 z+\alpha$.

\section{Solutions of the linearized problems}

Let us solve the linearized systems (\ref{small}) and
(\ref{small2}) and compare their solutions.

\noindent From the third equation and the forth  equation in
(\ref{small}), we obtain that \be
v_{zz}+\delta^2v_{xx}=0\label{17'}\ee Applying the method of
separation of variables, we seek the  solution of the equation
(\ref{17'}) in the form \be v(x,z,t)=F(x,t)G(z,t) \label{19}\ee
Substituting (\ref{19}) into the equation (\ref{17'}), separating
the variables and taking into account the expressions of $v$ on
the boundaries, that is, the fifth equation and the last equation
in (\ref{small}), we find \be v(x,z,t) =\f{1}{\sinh(K\delta)}\sinh
(K\delta z)\eta_t\label{22}\ee where $K\geq 0 $ is a constant that
might depend on time. Taking into account (\ref{22}) and the
fourth equation of the system (\ref{small}), we obtain \be
u(x,z,t)=\f{\delta}{K\sinh (K\delta)}\cosh (K\delta
z)\eta_{tx}+\omega_0 z+\mathcal{F}(x,t) \label{27}\ee where
$\mathcal{F}(x,t)$ is an arbitrary function. The components $u$
and $v$ of the velocity have to fulfill also the third equation in
(\ref{small}), hence, in view of (\ref{22}) and (\ref{27}), we get
\be \f{\delta}{K\sinh (K\delta)}\cosh (K\delta
z)\eta_{txx}+\f{\pa\mathcal{F}(x,t)}{\pa x}=-\f{K\delta }{\sinh
(K\delta)}\cosh (K\delta z)\eta_{t} \label{23'}\ee The above
relation must hold for all values of $x\in\mathbf{R}$, and $0\leq
z\leq 1$, thus, it follows \be \f{\pa\mathcal{F}(x,t)}{\pa
x}=0\label{24}\ee and \be \eta_{txx}+K^2\eta_t=0\label{25}\,\,
\footnote{For constants $K$ independent on time, we can integrate
the equation (\ref{25}) with respect to $t$ and we get \be
\eta_{xx}+K^2\eta=R(x)\nonumber \ee The solution of this equation
can be written into the form \ba \eta(x,t)=&&T(t)\left[c_1\cos(K
x)+c_2\sin(K x)\right]+\left[-\f 1{K}\int R(x)\sin(Kx)\,
dx\right]\cos(Kx)+\nonumber\\
&&\left[\f 1{K}\int R(x)\cos(Kx)\, dx\right]\sin(Kx) \nonumber\ea
$c_1$, $c_2$ being integration constants. We observe that to the
linearized problem (\ref{small}) we can get solutions different
from the usual one
 $\eta(x,t)=\cos(K(x-ct))$.}
\ee

\noindent Seeking periodic travelling wave solutions for the
equation (\ref{25}), we take \be K=2\pi\ee
  and we choose the following solution \be
\eta(x,t)=\cos(2\pi(x-ct)) \label{26}\ee where $c$ represents the
non-dimensional speed of propagation of the linear wave and is to
be determined. \\
 From (\ref{24}) the function
$\mathcal{F}(x,t)$ is independent of $x$, therefore we will denote
this function by $\mathcal{F}(t)$. Thus, the components of the
velocity filed are \ba &&u(x,z,t)=\f{2\pi\delta c}{\sinh
(2\pi\delta)}\cosh (2\pi\delta
z)\cos(2\pi(x-ct))+\omega_0z+\mathcal{F}(t)\nonumber\\
&&v(x,z,t)=\f{2\pi c}{\sinh(2\pi\delta)}\sinh (2\pi\delta
z)\sin(2\pi(x-ct)) \label{28'}\ea We return now to the systems
(\ref{small}) in order to find the the expressions of the
pressure. Taking into account the first  two equations in
(\ref{small}) and the expressions (\ref{28'}) of the velocity
field, we obtain  \be p(x,z,t)=\f{2\pi\delta
c^2}{\sinh(2\pi\delta)}\cosh(2\pi\delta z)\cos(2\pi(x-ct))-
x\mathcal{F}'(t) \label{29}\ee  On the free surface $z=1$ the
pressure (\ref{29}) has to fulfill the sixth equation of the
system (\ref{small}). Hence, in view of (\ref{26}), we get \be
2\pi\delta c^2\coth(2\pi\delta)\cos(2\pi(x-ct))-
x\mathcal{F}'(t)=\cos(2\pi(x-ct)) \ee The above relation must hold
for all values $x\in \mathbf{R}$, therefore, we get \be
\mathcal{F}(t)=\textrm{constant}:=c_0 \ee and we provide the
non-dimensional speed of the linear wave \be
c^2=\f{\tanh(2\pi\delta)}{2\pi\delta} \label{c}\ee Summing up, the
solution of the linear system (\ref{small}) is:
 \be
 \begin{array}{llll}
  \eta(x,t)=\cos(2\pi(x-ct))\\\cr
 p(x,z,t)=\f{2\pi\delta
c^2}{\sinh(2\pi\delta)}\cosh(2\pi\delta z)\cos(2\pi(x-ct))\\\cr
u(x,z,t)=\f{2\pi\delta c}{\sinh (2\pi\delta)}\cosh (2\pi\delta
z)\cos(2\pi(x-ct))+\omega_0z+c_0\\\cr v(x,z,t)=\f{2\pi
c}{\sinh(2\pi\delta)}\sinh (2\pi\delta z)\sin(2\pi(x-ct))
  \end{array}\label{solrotconst}\ee with $c$ given by (\ref{c}).

Taking into account (\ref{nondim}), (\ref{nodim'}),
(\ref{scaling}), (\ref{scaling'}), we return to the original
physical variables. The speed of the wave (\ref{c}) and the
solution (\ref{solrotconst}) become: \be
c=\pm\sqrt{gh_0}\sqrt{\f{\tanh(kh_0)}{kh_0}}=\pm\sqrt{g\f{\tanh(kh_0)}{k}}\label{c1}
\ee \be
 \hspace{0cm}\begin{array}{llll}
  \eta(x,t)=a\cos\left[(2\pi\left(\f{x}{\lambda}-\sqrt{\f{\tanh(kh_0)}{kh_0}}\f{\sqrt{gh_0}}{\lambda}t)\right)\right]=
  \epsilon h_0\cos[k(x-ct)]\\\cr
 p(x,z,t)=p_0+g(h_0-z)+\epsilon \f{g h_0}{\cosh(kh_0)}\cosh(k z)\cos[k(x-ct)]\\\cr
u(x,z,t)=\epsilon\f{kh_0c}{\sinh(kh_0)}\cosh (k
z)\cos[k(x-ct)]+\epsilon \omega_0\,z+\epsilon\sqrt{gh_0} c_0\\\cr
v(x,z,t)=\epsilon \f{kh_0c}{\sinh(kh_0)}\sinh (k z)\sin[k(x-ct)]
  \end{array}\label{solrotconst'}\ee
where \be k:=\f{2\pi}{\lambda}\ee is the wave number. The sign
minus in (\ref{c1}) indicates a left-going wave.

Let us look now at the linearized system (\ref{small2}).
 From the third equation and the forth  equation in
(\ref{small2}), we obtain again the equation (\ref{17'}). Applying
the method of separation of variables, we seek the solution of
this equation  in the form (\ref{19}). Substituting (\ref{19})
into the equation (\ref{17'}), separating the variables and taking
into account the expressions of $v$ on the boundaries, that is,
the fifth equation and the last equation in (\ref{small2}), we
find \be v(x,z,t) =\f{1}{\sinh(K\delta)}\sinh (K\delta
z)\left[\eta_t+(\omega_0+\alpha)\eta_x\right]\label{22'}\ee where
$K\geq 0 $ is a constant that might depend on time. Taking into
account (\ref{22'}) and the fourth equation of the system
(\ref{small2}), we obtain \be u(x,z,t)=\f{\delta}{K\sinh
(K\delta)}\cosh (K\delta
z)\left[\eta_{tx}+(\omega_0+\alpha)\eta_{xx}\right]+\mathfrak{F}(x,t)
\label{27'}\ee where $\mathfrak{F}(x,t)$ is an arbitrary function.
The components $u$ and $v$ of the velocity have to fulfill also
the third equation in (\ref{small2}), hence, in view of
(\ref{22'}) and (\ref{27'}), we get \ba &&\f{\delta}{K\sinh
(K\delta)}\cosh (K\delta
z)\left[\eta_{txx}+(\omega_0+\alpha)\eta_{xxx}\right]+\f{\pa\mathfrak{F}(x,t)}{\pa
x}=\nonumber\\
&&-\f{K\delta }{\sinh (K\delta)}\cosh (K\delta
z)\left[\eta_{t}+(\omega_0+\alpha)\eta_x\right] \label{23''}\ea
The above relation must hold for all values of $x\in\mathbf{R}$,
and $0\leq z\leq 1$, thus, it follows \be
\f{\pa\mathfrak{F}(x,t)}{\pa x}=0\label{24'}\ee and \be
\left[\eta_t+(\omega_0+\alpha)\eta_{x}\right]_{xx}+K^2\left[\eta_t+(\omega_0+\alpha)
\eta_{x}\right]=0\label{25'}\ee \noindent Seeking periodic
travelling wave solutions for the equation (\ref{25'}),  we take
\be K=2\pi\ee
 and we choose the following solution \be
\eta(x,t)=\cos(2\pi(x-ct)) \label{26'}\ee where $c$ represents the
non-dimensional speed of propagation of the linear wave and is to
be determined. \\
 From (\ref{24'}) the function
$\mathfrak{F}(x,t)$ is independent of $x$, therefore we will
denote this function by $\mathfrak{F}(t)$. Thus, the components
(\ref{27'}), (\ref{22'}) of the velocity filed are \ba
&&u(x,z,t)=\f{2\pi\delta\left( c-\omega_0-\alpha\right)}{\sinh
(2\pi\delta)}\cosh (2\pi\delta
z)\cos(2\pi(x-ct))+\mathfrak{F}(t)\nonumber\\
&&v(x,z,t)=\f{2\pi
\left(c-\omega_0-\alpha\right)}{\sinh(2\pi\delta)}\sinh
(2\pi\delta z)\sin(2\pi(x-ct)) \label{28''}\ea We return now to
the systems (\ref{small2}) in order to find the the expressions of
the pressure. Taking into account the first  two equations in
(\ref{small2}) and the expressions (\ref{28''}) of the velocity
field, we obtain  \ba p(x,z,t)&=&\f{2\pi\delta
(c-\omega_0-\alpha)}{\sinh(2\pi\delta)}(c-\omega_0
z-\alpha)\cosh(2\pi\delta
z)\cos(2\pi(x-ct))+\nonumber\\
&&+\f{\omega_0
(c-\omega_0-\alpha)}{\sinh(2\pi\delta)}\sinh(2\pi\delta
z)\cos(2\pi(x-ct))- x\mathfrak{F}'(t) \label{29'}\ea  On the free
surface $z=1$ the pressure (\ref{29'}) has to fulfill the sixth
equation of the system (\ref{small2}). Hence, in view of
(\ref{26'}), we get \ba
&&\left(c-\omega_0-\alpha\right)[2\pi\delta
(c-\omega_0-\alpha)\coth(2\pi\delta)+\omega_0]\cos(2\pi(x-ct))-\nonumber\\
&&\quad \quad \quad \quad\hspace{1cm} - x\mathfrak{F}'(t)=
\cos(2\pi(x-ct)) \ea The above relation must hold for all values
$x\in \mathbf{R}$, therefore, we get \be
\mathfrak{F}(t)=\textrm{constant}:=\mathfrak{c}_0 \ee and the
non-dimensional speed of the linear wave $c$ satisfies the
relation \be \left(c-\omega_0-\alpha\right)[2\pi\delta
(c-\omega_0-\alpha)\coth(2\pi\delta)+\omega_0]=1 \label{c'}\ee
Solving this equation we find \be
c=\omega_0+\alpha+\f{-\omega_0\pm\sqrt{\omega_0^2+8\pi\delta\coth(2\pi\delta)}}{4\pi\delta\coth(2\pi\delta)}
\label{c''}\ee Summing up, the solution of the linear system
(\ref{small2}) is given by   {\footnotesize\be \hspace{-0.5cm}
\begin{array}{llll}
  \eta(x,t)=\cos(2\pi(x-ct))\\\cr
 p(x,z,t)= \f{
(c-\omega_0-\alpha)}{\sinh(2\pi\delta)}\left[2\pi\delta(c-\omega_0
z-\alpha)\cosh(2\pi\delta z)+\omega_0 \sinh(2\pi\delta
z)\right]\cos(2\pi(x-ct))\\\cr u(x,z,t)=\f{2\pi\delta\left(
c-\omega_0-\alpha\right)}{\sinh (2\pi\delta)}\cosh (2\pi\delta
z)\cos(2\pi(x-ct))+\mathfrak{c}_0\\\cr v(x,z,t)=\f{2\pi
\left(c-\omega_0-\alpha\right)}{\sinh(2\pi\delta)}\sinh
(2\pi\delta z)\sin(2\pi(x-ct))
  \end{array}\label{solrotconst2}\ee} with $c$ from (\ref{c''}).

   Taking into account (\ref{nondim}),
(\ref{nodim'}), (\ref{scaling2}), we return to the original
physical variables. The speed of the wave (\ref{c''}) and the
solution (\ref{solrotconst2}) have in physical variables the
following expressions:
 {\footnotesize \ba c&=&\sqrt{gh_0}\left[\f{h_0}{\sqrt{gh_0}}\omega_0+
 \alpha+\f{-\f{h_0}{\sqrt{gh_0}}\omega_0\pm\sqrt{\f{h_0^2}{gh_0}\omega_0^2+4kh_0\coth(kh_0)}}{2kh_0
 \coth(kh_0)}\right]\nonumber\\
&=& h_0\omega_0+\alpha\sqrt{gh_0}+\f{1}{2k}\left[
-\omega_0\tanh(kh_0)\pm\sqrt{\omega_0^2\tanh^2(kh_0)+4gk\tanh(kh_0)}\right]
\label{cphys}\ea}
 {\footnotesize\be
 \hspace{-0.3cm}\begin{array}{lllll}
\eta(x,t)=
  \epsilon h_0\cos[k(x-ct)]\\\cr
 p(x,z,t)=p_0+g(h_0-z)+\epsilon \f{
\left(c-h_0\omega_0-\sqrt{gh_0}\alpha
\right)}{\sinh(kh_0)}\left[kh_0\left(c-\omega_0
z-\sqrt{gh_0}\alpha \right)\cosh(kz)\right.\\\cr
\hspace{2.5cm}+\left. h_0\omega_0 \sinh(k
z)\right]\cos[k(x-ct)]\\\cr
u(x,z,t)=\epsilon\f{kh_0\left(c-h_0\omega_0-\sqrt{gh_0}\alpha
\right)}{\sinh (kh_0)}\cosh (k z)\cos[k(x-ct)]+\omega_0
z+\alpha\sqrt{gh_0}+\epsilon\sqrt{gh_0}\mathfrak{c}_0\\\cr
v(x,z,t)=\epsilon\f{kh_0\left(c-h_0\omega_0-\sqrt{gh_0}\alpha
\right)}{\sinh (kh_0)}\sinh (k z)\sin[k(x-ct)]
  \end{array}\label{solrotconst2'}\ee}
where  \be k:=\f{2\pi }{\lambda}\ee is the wave number. The
solution
(\ref{cphys}), (\ref{solrotconst2'}) with $\alpha=0$, $\mathfrak{c}_0=0$ was also obtained in \cite{ev}.\\
Comparing (\ref{c1}), (\ref{solrotconst'}) with (\ref{cphys}),
(\ref{solrotconst2'}) we observe that \textit{the speed of the
wave and the pressure have different expressions in the two
linearizations. The velocity field has in the two linearizations
the form: \be
\begin{array}{ll}
u(x,z,t)=A\cosh (k z)\cos[k(x-ct)]+B z+C \\\cr
 v(x,z,t)=A\sinh (k
z)\sin[k(x-ct)]
\end{array}\label{uv}\ee
where, for the linearization (\ref{small}): \be
\begin{array}{ll}
c=\pm\sqrt{g\f{\tanh(kh_0)}{k}} \\\cr
A=\epsilon\f{kh_0c}{\sinh(kh_0)},\quad B=\epsilon \omega_0,\quad
C=\epsilon\sqrt{gh_0} c_0
\end{array}\label{abc} \ee and for
the linearization (\ref{small2}): {\footnotesize
\be\begin{array}{ll}
c-h_0\omega_0-\alpha\sqrt{gh_0}=\f{1}{2k}\left[
-\omega_0\tanh(kh_0)\pm\sqrt{\omega_0^2\tanh^2(kh_0)+4gk\tanh(kh_0)}\right]\\\cr
A=\epsilon\f{kh_0\left(c-h_0\omega_0-\sqrt{gh_0}\alpha
\right)}{\sinh (kh_0)},\quad B=\omega_0,\quad
C=\alpha\sqrt{gh_0}+\epsilon\sqrt{gh_0}\mathfrak{c}_0
\end{array}
\label{abc2}\ee}}

\section{Particle trajectories}

Let $\left(x(t), z(t)\right)$ be the path of a particle in the
fluid domain, with location $\left(x(0), z(0)\right):=(x_0,z_0)$
at time $t=0$. The motion of the particles  below the
 small-amplitude  water waves in constant
vorticity flows with the velocity field (\ref{uv}), is described
by the following differential system
  \be\left\{\begin{array}{ll}
 \f{dx}{dt}=u(x,z,t)=A\cosh (k
z)\cos[k(x-ct)]+Bz+C\\
\\
 \f{dz}{dt}=v(x,z,t)=A\sinh (kz)\sin[k(x-ct)]
 \end{array}\right.\label{diff2}\ee

The values of $A$, $B$, $C$ are either (\ref{abc}) or
(\ref{abc2}), depending on which linearization we consider. From
(\ref{abc}), (\ref{abc2}), for any $\omega_0\neq 0$, we get  in
the both cases
 \be A\neq 0\label{a}\ee
\\

\textit{For the first linearization (\ref{small}), the sign of $A$
depends  on the sign of the wave speed $c$. Thus, if we choose in
(\ref{abc}) the square root with minus, that is, we consider
left-going waves, we have $A<0$ and if we choose in (\ref{abc})
the square root with plus, that is, we consider right-going waves,
we get $A>0$.}

 \textit{For the
second linearization (\ref{small2}), the sign of $A$ depends on
the sign of  $c-h_0\omega_0-\sqrt{gh_0}\alpha$. Looking at the
expression (\ref{abc2}) of $c-h_0\omega_0-\sqrt{gh_0}\alpha$, we
get  that \textbf{independent of the sign of $\omega_0$}, if we
choose in (\ref{abc2}) the square root with minus then
$c-h_0\omega_0-\sqrt{gh_0}\alpha<0$, thus, $A<0$, and if we choose
in (\ref{abc2}) the square root with plus then
$c-h_0\omega_0-\sqrt{gh_0}\alpha>0$, thus, $A>0$. The expression
$c-h_0\omega_0-\sqrt{gh_0}\alpha$ could be regarded as "the speed"
of a wave, which is left-going if we take the square root with
minus in (\ref{abc2}), and is right-going if we take the square
root with plus in (\ref{abc2}).}
\\

\noindent Indeed,  if we choose in (\ref{abc2}) the square root
with minus then, how $k$ and $h_0$ are greater then zero, for
$\omega_0>0$, the expression $
-\omega_0\tanh(kh_0)-\\
\sqrt{\omega_0^2\tanh^2(kh_0)+4gk\tanh(kh_0)}$ is evidently
smaller then zero, and for \\
$\omega_0<0$,
$-\omega_0\tanh(kh_0)-\sqrt{\omega_0^2\tanh^2(kh_0)+4gk\tanh(kh_0)}<0$
is equivalent with
$-\omega_0\tanh(kh_0)<\sqrt{\omega_0^2\tanh^2(kh_0)+4gk\tanh(kh_0)}$.
By raising to the power two, the last inequality is equivalent
with $\tanh{kh_0}>0$, which is a true inequality for $k$ and $h_0$
are greater then zero.$\square$\\

To study the exact solution of the system (\ref{diff2}) it is
 more convenient to re-write it in the following moving frame
 \be
 X=k(x-ct),\quad  Z=k z \label{frame}
 \ee
This transformation yields \be\left\{\begin{array}{ll}
 \f{dX}{dt}=kA\cosh(Z)\cos(X)+BZ +
 k(C-c)\\
 \\
 \f{dZ}{dt}=kA \sinh(Z)\sin(X)
 \end{array}\right.\label{diff3}\ee
We write the second equation of this system in the form \be
\f{dZ}{\sinh(Z)}=kA\sin X(t)\,dt \label{44}\ee Integrating, we get
\be \log\left[\tanh\left(\f Z{2}\right)\right]=\int kA \sin
X(t)\,dt\label{X} \ee If \be \int kA \sin X(t)\,dt<0
\label{int}\ee then \be Z(t)=2\textrm{ arctanh
}\left[\exp\left(\int kA \sin X(t)\,dt\right)\right] \label{42}\ee
Taking into account the formula: \be
\cosh(2x)=\f{1+\tanh^2(x)}{1-\tanh^2 (x)}, \label{45}\ee and the
expression (\ref{42}) of $Z(t)$, the first equation of the system
(\ref{diff3}) becomes
 \be \f{d
X}{dt}=kA\f{1+w^2}{1-w^2}\cos (X)+2B \textrm{ arctanh
}(w)+k(C-c)\label{35'}\ee where we have denoted by \be
w=w(t):=\exp\left(\int k A \sin X(t)\,dt\right) \label{31}\ee With
(\ref{int}) in view, we have \be 0<w<1\label{w} \ee
 From (\ref{31}) we get \be
kA\sin X(t)=\f 1{w(t)}\f{d w}{dt} \label{32}\ee Differentiating
with respect to $t$ this relation, we obtain
 \be k A\cos (X)\f{dX}{dt}=\f 1{w^2}\left[\f{d^2w}{dt^2}w-\left(\f{dw}{dt}
 \right)^2\right]\label{33}
\ee From (\ref{32}) we have furthermore \be k^2A^2\cos^2
(X)=k^2A^2-\f 1{w^2}\left(\f{d w}{dt}\right)^2\label{34} \ee Thus,
taking into account (\ref{33}), (\ref{34}), the equation
(\ref{35'}) becomes
\begin{eqnarray} &&\f{d^2w}{dt^2}+\f{2w}{1-w^2}\left(\f{dw}{dt}
\right)^2-k^2A^2 w\f{1+w^2}{1-w^2}-\nonumber\\
 &&\hspace{1cm}- \sqrt{k^2A^2w^2-\left(\f{dw}{dt}
 \right)^2}\Big[2B\textrm{ arctanh }(w)+k(C-c)\Big]=0\label{36'}
\end{eqnarray}
We make the following substitution
 \be \xi^2(w):=k^2A^2w^2-\left(\f{dw}{dt}\right)^2 \label{37}\ee
$A$ being different from zero (\ref{a}).
 Differentiating with respect to $t$ this relation, we get
 \be
\xi\f{d\xi}{dw}=k^2A^2w-\f{d^2w}{dt^2}
 \label{38}\ee
We replace (\ref{37}), (\ref{38}) into the equation (\ref{36'})
and we obtain  the equation \be
\xi\f{d\xi}{dw}+\f{2w}{1-w^2}\xi^2+\Big[2B\textrm{ arctanh
}(w)+k(C-c)\Big]\xi=0 \label{39c}\ee A solution of the equation
(\ref{39c}) is \be \xi=0\ee which, in view of (\ref{37}) and
(\ref{32}) implies \be \sin X(t)=\pm 1 \ee Therefore, from
(\ref{42}) with the condition (\ref{int}),  and further from
(\ref{frame}), \textit{a solution of the system (\ref{diff2})  is
\be
\begin{array}{ll}
 x(t)=ct+\textrm{const}_1\\
 \\
 z(t)=\f 2{k}\textrm{ arctanh }\left[\exp\left(- |kA\,t+\textrm{const}_2|\right)\right]
  \end{array} \label{sol0}\ee}
$const_1$ and $const_2$ are constants determined by the initial
conditions $\left(x(0), z(0)\right):=(x_0,z_0)$. This peakon-like
solution was also presented in the paper \cite{io5}. The graph of
the parametric curve (\ref{sol0}) is drawn  in the
  Figure 2.

\vspace{0.6cm}

\hspace{0cm}\scalebox{0.40}{\includegraphics{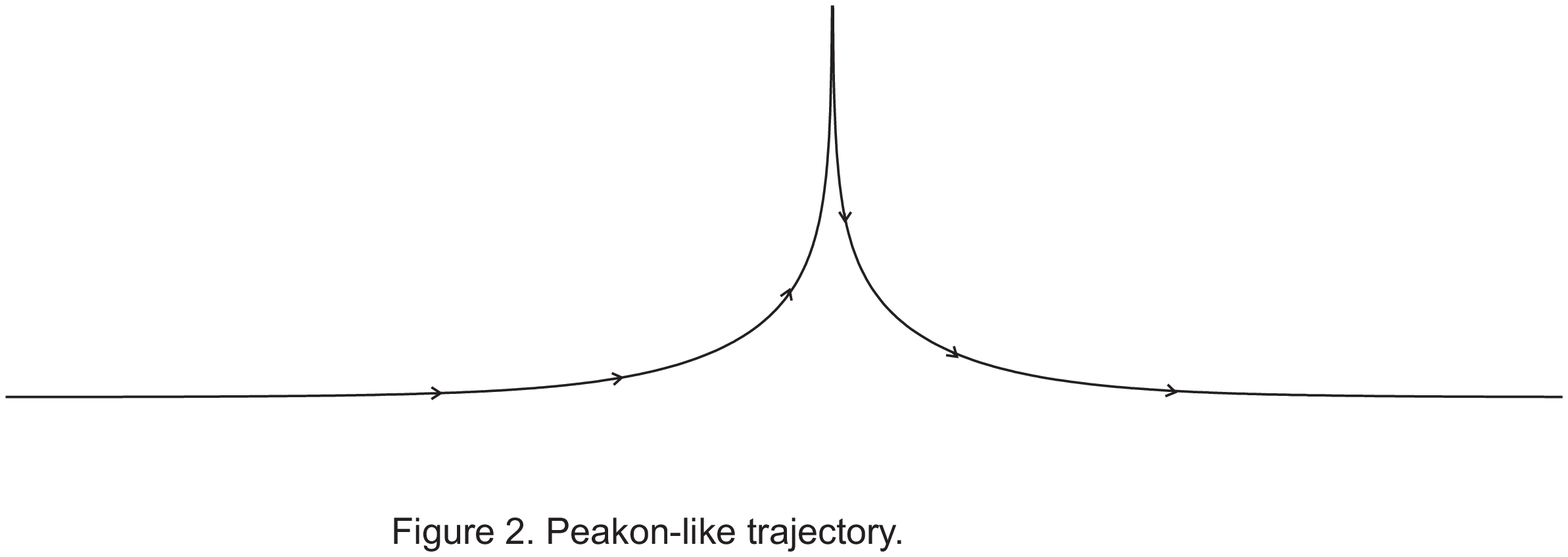}}
\\

\noindent Calculating the derivatives of $x(t)$ and $z(t)$ with
respect to $t$, we get
 \be
\begin{array}{ll}
x'(t)=c\\
\cr z'(t)=\left\{\begin{array}{c} -\f{2A\exp\left[-(kA\,t+const_2)
\right]}{1-\exp\left[- 2(kA\,t+const_2)\right]},\, \,
 kA\,t+const_2>0\\
\f{2A\exp\left[kA\,t+const_2\right]}{1-\exp\left[2(kA\,t+const_2)\right]}
,\, \, kA\,t+const_2<0\end{array}\right.
\end{array}\ee
Hence, \textit{for the solution (\ref{sol0}) a stagnation point in
the fluid,  where  $x'(t)=c$, $z'(t)=0$, appear only for
$t\rightarrow \pm\infty$. We observe that at these points the path
of the particle has a horizontal tangent}.

The other solutions of the equation (\ref{39c}) satisfy
 \be
\f{d\xi}{dw}+\f{2w}{1-w^2}\xi=-\Big[2B\textrm{ arctanh
}(w)+k(C-c)\Big] \label{39c'} \ee The homogeneous equation: \be
\f{d\xi}{dw}+\f{2w}{1-w^2}\xi=0\ee has the solution \be \xi(w)=
\mathcal{\theta}(1-w^2)\ee where $\theta$ is an integration
constant. By the method of variation of constants, the general
solution of the non-homogeneous equation (\ref{39c'}) is given by
\be \xi(w)=\theta(w)(1-w^2) \ee where $\theta(w)$ is a continuous
function which satisfies the equation \be
\f{d\theta}{dw}=-\f{1}{1-w^2}\Big[2B\textrm{ arctanh
}(w)+k(C-c)\Big] \label{40c}\ee The solution of the equation
(\ref{40c}) is \be \theta(w)=-B\textrm{ arctanh
}^2(w)-k(C-c)\textrm{ arctanh }(w)+\beta\ee $\beta$ being a
constant. Therefore, the solution of the non-homogeneous equation
(\ref{39c'}) has the expression \be
\xi(w)=(1-w^2)\left[\beta-k(C-c)\textrm{ arctanh }(w)-B\textrm{
arctanh }^2(w)\right] \ee Taking into account (\ref{37}), we get
{\footnotesize\be
\f{dw}{dt}=\pm\sqrt{k^2A^2w^2-(1-w^2)^2\left[\beta-k(C-c)\textrm{
arctanh }(w)-B\textrm{ arctanh }^2(w)\right]^2}\label{eq} \ee}
We separate the variables in (\ref{eq}):{\footnotesize\be
\pm\f{dw}{(1-w^2)\sqrt{k^2A^2\f{w^2}{(1-w^2)^2}-\left[\beta-k(C-c)\textrm{
arctanh }(w)-B\textrm{ arctanh }^2(w)\right]^2}}=dt\label{41} \ee}
From (\ref{42}), (\ref{31}), we have \be \f{Z(t)}{2}=\textrm{
arctanh }(w)\label{zeta}\ee Thus, (\ref{41}) can be written as \be
\pm\f{dZ}{2\sqrt{k^2A^2\f{\tanh^2(\f{Z}{2})}{(1-
\tanh^2(\f{Z}{2}))^2}-\left[\beta-\f{k(C-c)}{2}Z-\f{B}{4}
Z^2\right]^2}}=dt \ee that is,\be
\pm\f{dZ}{\sqrt{k^2A^2\sinh^2(Z)-\left[2\beta-k(C-c)Z-\f{B}{2}
Z^2\right]^2}}=dt\label{46} \ee By (\ref{44}) we  obtain \be
X(t)=\arcsin \left[\f 1{kA}\f 1{\sinh(Z(t))}\f{dZ(t)}{dt}\right]
\ee Further, from (\ref{frame}), we get \textit{another solution
of the system (\ref{diff2}): \be
\begin{array}{lll}
 x(t)= ct+\f{1}{k}\arcsin \left[\f 1{kA}\f
1{\sinh(Z(t))}\f{dZ(t)}{dt}\right]\\
\\
 z(t)=\f 1{k} Z(t)
  \end{array} \label{sol2}\ee
 $Z(t)$ being the solution of the equation (\ref{46})}.
\\

We observe that the solutions (\ref{sol2}) \textit{are not closed
curves}.
\\

\noindent Indeed, if there exists $t_2>t_1$ such that
$Z(t_2)=Z(t_1)$, then, in view of (\ref{46}), we also have
$\f{dZ}{dt}(t_1)=\pm\sqrt{k^2A^2\sinh^2(Z(t_1))-\left[2\beta-k(C-c)Z(t_1)-\f{B}{2}
Z^2(t_1)\right]^2}$\\
$=\pm\sqrt{k^2A^2\sinh^2(Z(t_2))-\left[2\beta-k(C-c)Z(t_2)-\f{B}{2}
Z^2(t_2)\right]^2}=\f{dZ}{dt}(t_2)$. Thus, although in the moving
frame we obtain in this case a closed curve with $Z(t_1)=Z(t_2)$
and $X(t_1)=X(t_2)$, in the fixed frame  we get
 $z(t_2)=z(t_1)$ and
$x(t_2)-x(t_1)
=c(t_2-t_1)\neq 0$. If $c>0$, the particles which follow these
curves will have a \textit{forward drift}, if $c<0$, they will
have a \textit{backward drift}.$\square$
\\

Let us now investigate more the equation (\ref{46}). Using the
formula: $\sinh^2(x)=\f{\cosh(2x)-1}{2}$, the equation (\ref{46})
can be written in the form:
 \be
\pm\f{dZ}{\sqrt{\f{k^2A^2}{2}\cosh(2Z)-\f{k^2A^2}{2}-\left[2\beta-k(C-c)Z-\f{B}{2}
Z^2\right]^2}}=dt \label{43}\ee Taking into account the expression
of  $\cosh(x)$ as Taylor series: \be
\cosh(x)=1+\f{x^2}{2!}+\f{x^4}{4!}+\f{x^6}{6!}+\cdots=\sum^\infty_{n=0}\f{x^{2n}}{(2n)!}
\label{cosh}\ee we get  under the square root in (\ref{43}) the
following power series \ba &&-4\beta^2+4k(C-c)\beta Z
+\left[k^2A^2+2\beta B-k^2(C-c)^2\right]
Z^2-Bk(C-c)Z^3+\nonumber\\
&&\hspace{2cm} +\left(\f {k^2A^2}{3}-\f{B^2}{4}\right)
Z^4+\f{2^5k^2A^2}{6!}Z^6+\f{2^7k^2A^2}{8!}Z^8+\cdots \label{49}\ea
The constant $A$ is different from zero (\ref{a}), thus,  the
power series (\ref{49}) contains for sure powers of $Z$ higher
than four. A partial sum of the above series is a polynomial of
degree higher than four. Thus, considering only a partial sum of
this series,
 the solution of the equation
(\ref{46}) involves a \textit{hyperelliptic integral} (for
hyperelliptic integrals see, for example, \cite{byrd}, page 252).
Its inversion would lead to a \textit{hyperelliptic function}.

There are special cases when a hyperelliptic integral can be
reduced to an elliptic one and thus,  its inversion will contain
the Jacobi elliptic functions sn, cn, sc, etc.
 If
in (\ref{49}) we have $C=c$ (by choosing appropriate constants
$c_0$, $\mathfrak{c}_0$  in (\ref{abc}), respectively
(\ref{abc2})) and we consider powers of $Z$ till six, the solution
of the equation (\ref{46}) involves the following hyperelliptic
integral \be \pm\int\f{dZ}{\sqrt{\f{2k^2A^2}{45}Z^6+\left(\f
{k^2A^2}{3}-\f{B^2}{4}\right) Z^4+\left(k^2A^2+2\beta B\right)
Z^2-4\beta^2}}=t\label{46'} \ee We consider the substitution \be
Z^2=\f 1{\hat{Z}}  \label{hat}\ee and thus, the left-hand side in
 (\ref{46'}) reduces to an elliptic integral of the first kind:
 \be
 \pm\int\f{d\hat{Z}}{-2\sqrt{-4\beta^2\hat{Z}^3+\left(k^2A^2
 +2\beta B\right)
\hat{Z}^2+\left(\f {k^2A^2}{3}-\f{B^2}{4}\right)
\hat{Z}+\f{2k^2A^2}{45}}}=t
 \label{50}\ee
This elliptic integral of the first kind   may by reduced to the
Legendre normal form.

\textbf{Case 1}: all the zeroes of the cubic polynomial under the
square root in (\ref{50}) are real and distinct. We denote them by
$\hat{Z}_1<\hat{Z}_2<\hat{Z}_3$.  Because the leading coefficient
of this cubic polynomial  is smaller then zero and its constant
term is greater then zero, we have either \be
0<\hat{Z}_1<\hat{Z}_2<\hat{Z}_3
 \label{52}\ee
or \be \hat{Z}_1<\hat{Z}_2<0<\hat{Z}_3\label{53} \ee

\textbf{Case 1a}: the condition (\ref{52}) is fulfilled. \\
 Then
 we introduce the variable
$\varphi$ by (see \cite{smirnov} Ch. VI, \S 4, page 602) \be
\hat{Z}=\hat{Z}_2\sin^2\varphi+\hat{Z}_3\cos^2\varphi>0
\label{varphi}\ee and  we get \ba &&
-4\beta^2(\hat{Z}-\hat{Z}_1)(\hat{Z}-\hat{Z}_2)(\hat{Z}-\hat{Z}_3)=\nonumber\\
&&\hspace{0.5cm}=4\beta^2
\sin^2\varphi\cos^2\varphi(\hat{Z}_3-\hat{Z}_2)^2(\hat{Z}_3-\hat{Z}_1)\left(
1-k_1^2\sin^2\varphi\right)>0\nonumber\\
&&d\hat{Z}=-2\sin\varphi\cos\varphi(\hat{Z}_3-\hat{Z}_2)
d\varphi\nonumber\ea where the constant $0<k_1^2<1$ is given by
\be
k_1^2:=\f{\hat{Z}_3-\hat{Z}_2}{\hat{Z}_3-\hat{Z}_1}\label{k1}\ee
Therefore we obtain the Legendre normal form of the integral in
(\ref{50}):\be
\f{1}{\mathcal{C}_1}\int\f{d\varphi}{\sqrt{1-k_1^2\sin^2\varphi
}}=t \label{51}\ee the constant factor in front of the integral
being equal to\be
\mathcal{C}_1:=\pm2|\beta|\sqrt{\hat{Z}_3-\hat{Z}_1}\label{C}\ee
 The inverse of the integral in (\ref{51}) is sn (the Jacobi elliptic function sine amplitude, see, for example,
\cite{byrd})  \be \textrm{ sn }\left(\mathcal{C}_1\,t;k_1\right)
=\sin\varphi\label{56}\ee
 In view of the notations
(\ref{hat}) and (\ref{varphi})  we get \be
Z(t)=\f{1}{\sqrt{\hat{Z}_2\textrm{ sn
}^2\left(\mathcal{C}_1\,t;k_1\right)+\hat{Z}_3\textrm{ cn
}^2\left(\mathcal{C}_1\,t;k_1\right)}} \label{Z1}\ee
 cn being the
Jacobi elliptic function cosine amplitude (see, for example,
\cite{byrd}). Taking into account the expressions for the
derivatives of sn and cn (see, for example, \cite{byrd}), that is,
 \ba&&\f {d}{dt}\textrm{ sn }(t;k)=\textrm{ cn }(t;k)
 \textrm{ dn }(t;k)\nonumber\\
 &&\f {d}{dt}\textrm{ cn }(t;k)=-\textrm{ sn }(t;k)\textrm{ dn }(t;k),\nonumber\ea
where $\textrm{ dn }(t;k):=\sqrt{1-k^2\textrm{ sn }^2(t;k)}$, we
obtain \be
\f{dZ(t)}{dt}=\f{\mathcal{C}_1(\hat{Z}_3-\hat{Z}_2)\textrm{ sn
}\left(\mathcal{C}_1\,t;k_1\right)\textrm{ cn
}\left(\mathcal{C}_1\,t;k_1\right)\textrm{ dn
}\left(\mathcal{C}_1\,t;k_1\right)}{\left[\sqrt{\hat{Z}_2\textrm{
sn }^2\left(\mathcal{C}_1\,t;k_1\right)+\hat{Z}_3\textrm{ cn
}^2\left(\mathcal{C}_1\,t;k_1\right)}\right]^3}
 \label{Z1'}\ee
We introduce  (\ref{Z1}) and (\ref{Z1'}) in (\ref{sol2}) and we
get  $x(t)$ and $z(t)$ explicitly.

We remark that, if $\f {k^2A^2}{3}-\f{B^2}{4}<0$ and $
{k^2A^2}+2\beta B>0$ (this can happen, for example, for  a small
enough $A$ and for ($B>0$ $\&$ $\beta>0$) or ($B<0$ $\&$
$\beta<0$)), the coefficients of the cubic polynomial  in
(\ref{50}) have alternating signs. Thus, by Descartes' rule of
signs, if all the roots are real, the situation (\ref{52}) occurs.

\textbf{Case 1b}: the condition (\ref{53}) is fulfilled.\\
$\hat{Z}_1,\,\hat{Z}_2,\,\hat{Z}_3$ being the zeroes of the real
cubic polynomial under the square root in (\ref{50}), this
polynomial has the unique decomposition
$-4\beta^2(\hat{Z}-\hat{Z}_1)(\hat{Z}-\hat{Z}_2)(\hat{Z}-\hat{Z}_3)$.
A suitable change of variable transforms the integral (\ref{50})
to the Legendre normal form (\ref{51}) up to a constant. In
general, the elliptic functions can  have complex arguments (for
example, if the constant factor in front of the integral
(\ref{51}) is a complex number, then  the obtained sine amplitude
function will depend on  a complex variable) but here we are
interested only in the real case. We are also looking for a real
$Z$, so, $\hat{Z}$ introduced by (\ref{hat}) has to be greater
then zero.
 With the change of variable (\ref{varphi}), which brings the
integral (\ref{50}) to the Legendre normal form (\ref{51}),
because now $\hat{Z}_2<0$, we end up with a $\hat{Z}$ which can be
positive, negative or zero. Thus, in this case, we get the
expression (\ref{Z1}) of $Z(t)$ only if  $t$ satisfies \be
\hat{Z}_2\textrm{ sn
}^2\left(\mathcal{C}_1\,t;k_1\right)+\hat{Z}_3\textrm{ cn
}^2\left(\mathcal{C}_1\,t;k_1\right)>0 \label{57}\ee that is, by
$\textrm{ sn }^2\left(\mathcal{C}_1\,t;k_1\right)+\textrm{ cn
}^2\left(\mathcal{C}_1\,t;k_1\right)=1$, only if $t$  satisfies
\be \textrm{ sn
}^2\left(\mathcal{C}_1\,t;k_1\right)<\f{\hat{Z}_3}{\hat{Z}_3-\hat{Z}_2}
\label{58}\ee where $0<\f{\hat{Z}_3}{\hat{Z}_3-\hat{Z}_2}<1$.
\\
For a very small positive solution  $\hat{Z}_3\rightarrow 0$, the
set of $t$'s which fulfill the above inequality (\ref{58}) tends
to the empty set. Thus, in this case, the hyperelliptic integral
in (\ref{46'}) can not be reduced to an elliptic one and the
solution can \textit{not} be expressed with the aid of the Jacobi
elliptic functions. The solution will be expressed with the aid of
a hyperelliptic function  obtained by the inversion of the
integral in  (\ref{46'}).

\textbf{Case 2}: the cubic polynomial under the square root in
(\ref{50}) has only one real solution denoted $\hat{Z}_0$. Because
the leading coefficient of this cubic polynomial  is smaller then
zero and its constant term is greater then zero, we have \be
\hat{Z}_0>0 \ee
 We denote by $p$ and $q$ the real coefficients such that
{\footnotesize\be -4\beta^2\hat{Z}^3+\left(k^2A^2+2\beta B\right)
\hat{Z}^2+\left(\f {k^2A^2}{3}-\f{B^2}{4}\right)
\hat{Z}+\f{2k^2A^2}{45}=-4\beta^2(\hat{Z}-\hat{Z}_0)(\hat{Z}^2+p\hat{Z}+q)\label{60}\ee}
We introduce the variable $\psi$ by (see \cite{smirnov} Ch. VI, \S
4, page 602) \be
\hat{Z}=\hat{Z}_0-\sqrt{\hat{Z}_0^2+p\hat{Z}_0+q}\,\tan^2\f{\psi}{2}
\label{varphi2}\ee and we get \ba &&
-4\beta^2(\hat{Z}-\hat{Z}_0)(\hat{Z}^2+p\hat{Z}+q)=\nonumber\\
&&\hspace{1.5cm}=4\beta^2
\left(\sqrt{\hat{Z}_0^2+p\hat{Z}_0+q}\right)^3\f{\tan^2\f{\psi}{2}}{\cos^4\f{\psi}{2}}\left(
1-k_2^2\sin^2\varphi\right)>0\nonumber\\
&&d\hat{Z}=-\sqrt{\hat{Z}_0^2+p\hat{Z}_0+q}\f{\tan\f{\psi}{2}}{\cos^2\f{\psi}{2}}
d\psi\nonumber\ea where the constant $0<k_2^2<1$ is given by \be
k^2_2:=\f
1{2}\left(1+\f{\hat{Z}_0+\f{p}{2}}{\sqrt{\hat{Z}_0^2+p\hat{Z}_0+q}}\right)\label{k2}
\ee Therefore we obtain the Legendre normal form of the integral
in (\ref{50}): \be
\f{1}{\mathcal{C}_2}\int\f{d\psi}{\sqrt{1-k_2^2\sin^2\psi }}=t
\label{51'}\ee the constant factor in front of the integral being
equal to \be
\mathcal{C}_2:=\pm4|\beta|(\hat{Z}_0^2+p\hat{Z}_0+q)^\f1{4}\label{C2}\ee
 The inverse of the integral in (\ref{51'}) is
   \be \textrm{ sn }\left(\mathcal{C}_2\,t;k_2\right)
=\sin\psi\label{56'}\ee
 Taking into account
(\ref{varphi2}), we get \be
\hat{Z}(t)=\hat{Z}_0-\sqrt{\hat{Z}_0^2+p\hat{Z}_0+q}\,\f{1-\textrm{
cn }\left(\mathcal{C}_2\,t;k_2\right)}{1+\textrm{ cn
}\left(\mathcal{C}_2\,t;k_2\right)} \label{Z2}\ee  If  $t$
satisfies the following inequality \be \textrm{ cn
}\left(\mathcal{C}_2\,t;k_2\right)>\f{\sqrt{\hat{Z}_0^2+p\hat{Z}_0+q}-\hat{Z}_0}{\sqrt{\hat{Z}_0^2+p\hat{Z}_0+q}
+\hat{Z}_0}\label{59} \ee where
$-1<\f{\sqrt{\hat{Z}_0^2+p\hat{Z}_0+q}-\hat{Z}_0}{\sqrt{\hat{Z}_0^2+p\hat{Z}_0+q}
+\hat{Z}_0}<1$, then, $\hat{Z}$ from (\ref{Z2}) is greater than
zero  and we obtain  by (\ref{hat}) the expression of $Z(t)$.
 We can
calculate the time derivative of $Z(t)$ and by (\ref{sol2}) we get
$x(t)$ and $z(t)$ explicitly.

 As in the case 1b,
 for very small
positive solution  $\hat{Z}_0\rightarrow 0$,  the set of $t$'s
which fulfill the  inequality (\ref{59}) tends to the empty set.
Thus, in this case, the hyperelliptic integral in (\ref{46'}) can
not be reduced to an elliptic one and the solution can
\textit{not} be expressed with the aid of the Jacobi elliptic
functions. The solution will be expressed with the aid of a
hyperelliptic function  obtained by the inversion of the integral
in (\ref{46'}).


We remark that, if $\f {k^2A^2}{3}-\f{B^2}{4}<0$ and $
{k^2A^2}+2\beta B<0$ (this can happen, for example, for  a small
enough $A$ and for ($B<0$ $\&$ $\beta>0$) or ($B>0$ $\&$
$\beta<0$)), the coefficients of the cubic polynomial in
(\ref{50}) have signs - - - +. Thus, by Descartes' rule of signs,
only one root is positive and we are in the case 1b or in the case
2. This positive root it will be close to zero.
\\

Let us draw below some of the curves  obtained for different
values of the parameters, using  Mathematica\footnote{In
Mathematica the Jacobi elliptic functions  are implemented  as
JacobiSN$[u,m:=k^2_1]$:= sn$(u;k_1)$,
JacobiCN$[u,m:=k^2_1]$:=cn$(u;k_1)$,
JacobiDN$[u,m:=k^2_1]$:=dn$(u;k_1)$}.\\
We consider  $k=1$, $h_0=1$, $g=9.8$, $\epsilon=0.1$, $\alpha=0$,
$\beta=1$ and $\omega_0=2>0$. Then, by  (\ref{abc2}), choosing the
square root with sign plus, we get $c=4.07454>0$, $A=0.176526>0$,
$B=2>0$. We take $\mathfrak{c}_0$ such that $C=c$. In this case,
all the roots of the cubic polynomial under the square root in
(\ref{50}) are real and we get $Z(t)$ in the form (\ref{Z1}). The
graph of the curve obtained  is drawn in   Figure 3.
\\

\hspace{0.5cm} \scalebox{0.40}{\includegraphics{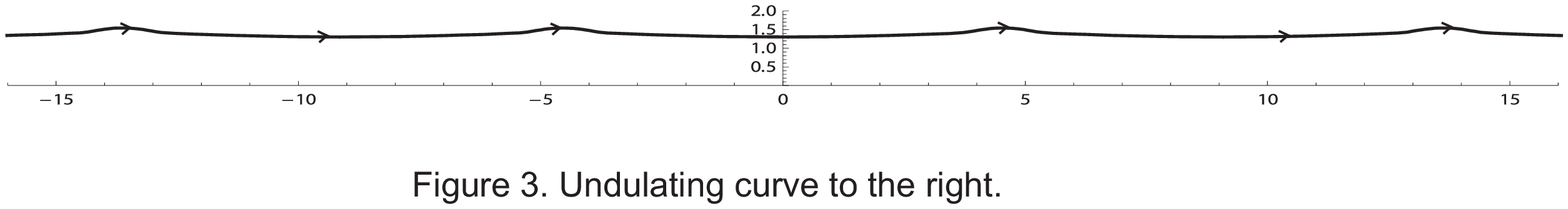}}
\\

\noindent Using the same values for $k$, $h_0$, $g$, $\epsilon$,
$\alpha$ and $\beta$ but taking $\omega_0=20>0$, we get by
(\ref{abc2}), choosing the square root with sign minus,
$c=4.29294>0$, $A=-1.33654<0$, $B=20>0$. We take $\mathfrak{c}_0$
such that $C=c$. Then all the roots of the cubic polynomial under
the square root in (\ref{50}) are real and we get $Z(t)$ in the
form (\ref{Z1}). The graph of the  curve obtained is depicted in
Figure 4.
\\

\hspace{0.5cm} \scalebox{0.40}{\includegraphics{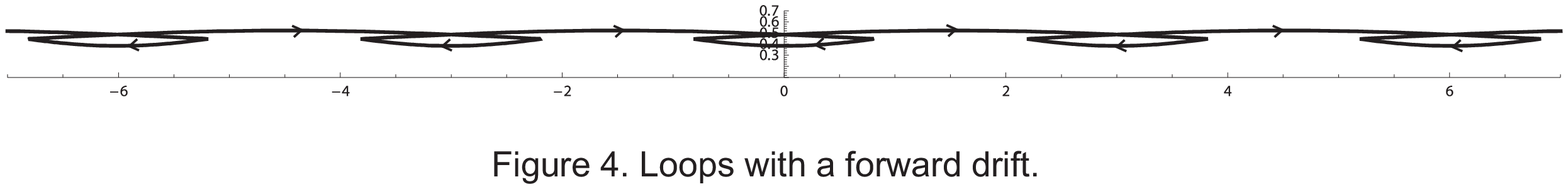}}
\\

\noindent For the same $k$, $h_0$, $g$,  $\epsilon$, $\alpha$ and
$\beta$ as above, with $\omega_0=2>0$ but choosing the square root
with sign minus  in (\ref{abc2}), we get $c=-1.59773<0$,
$A=-0.306137<0$, $B=2>0$. We take $\mathfrak{c}_0$ such that
$C=c$. Then all the roots of the cubic polynomial under the square
root in (\ref{50}) are real and we get $Z(t)$ in the form
(\ref{Z1}). The graph of the  curve obtained is presented in
Figure 5.
\\

\hspace{0.5cm} \scalebox{0.40}{\includegraphics{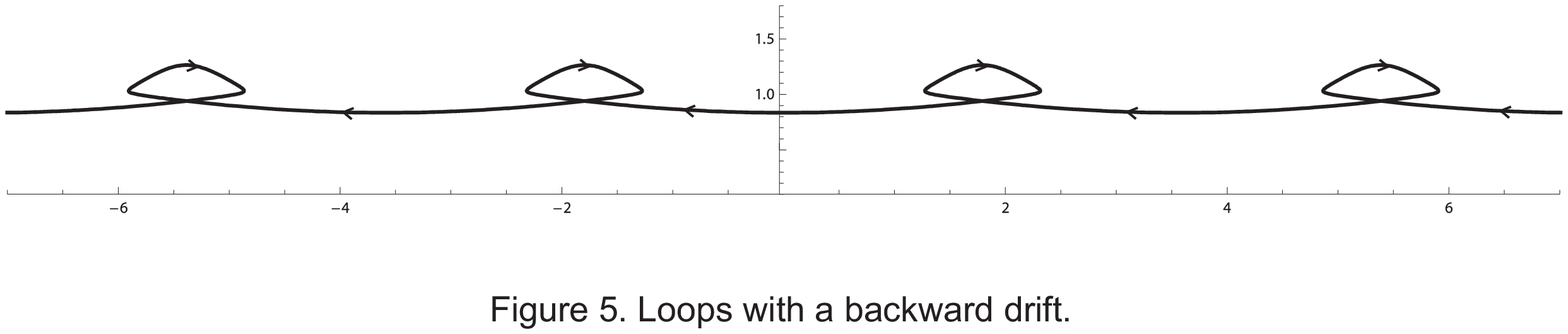}}
\\

\noindent We choose now $\omega_0=-20<0$, $k$, $h_0$, $g$,
$\epsilon$ $\alpha$ and $\beta$ having the same values as above.
We get by (\ref{abc2}), choosing the square root with sign plus,
$c=-4.29294<0$, $A=1.33654>0$, $B=-20<0$. We take $\mathfrak{c}_0$
such that $C=c$. Then the cubic polynomial under the square root
in (\ref{50}) has only one real root, that is,
$\hat{Z}_0=0.000798$. The real coefficients $p$ and $q$ from
(\ref{60}) have the values: $p=9.55422$, $q=24.8588$ and the right
hand side in (\ref{59}) has the value 0.99968. We conclude that in
this case the solution can not be expressed through Jacobi
elliptic functions.
 \\

 We
would like to make some remarks on the stagnation points inside
the fluid. Calculating the derivatives  with respect to $t$ of
$x(t)$ and $z(t)$ from (\ref{sol2}), we get
 \be
\begin{array}{ll}
x'(t)=c+\f1{k}\f{\sinh(Z)\f{d^2Z}{dt^2}-\cosh(Z)\left(\f{dZ}{dt}\right)^2}{\sinh(Z)\sqrt{k^2A^2\sinh^2
(Z)-\left(\f{dZ}{dt}\right)^2}}\\
\cr z'(t)=\f 1{k}\f{dZ}{dt}\end{array}\label{48} \ee where, taking
into account (\ref{46}),  \be \f{d^2Z}{dt^2}=
k^2A^2\sinh(Z)\cosh(Z)-[2\beta-k(C-c) Z-\f{B}{2}Z^2][-k(C-c)-B
Z]\label{Z} \ee With (\ref{46}) in view, for those $Z(t)$
satisfying the following equation \be \Big|
kA\sinh(Z)\Big|=\Big|2\beta-k(C-c) Z-\f{B}{2}Z^2\Big|
\label{47}\ee
we have \be \f{dZ}{dt}=0,\quad \f{d^2Z}{dt^2}=0,\ee
 and thus, $x'(t)$, $z'(t)$  from
 (\ref{48}) becomes
 \be
 x'(t)=c,\quad z'(t)=0
 \ee
Hence, \textit{for the solution (\ref{sol2}) the stagnation points
in the fluid are obtained by solving the equation (\ref{47})}.

The equation (\ref{47}) can be solved graphically. Depending on
the signs and on the values of the parameters A, B, C, c and
$\beta$, the equation (\ref{47})  can have one,  two, three, four
or six solutions.
 See, for
example, in Figure 6 some possibilities that can occur. With
continuous line we have drawn $\Big| kA\sinh(Z)\Big|$.
Which of these solutions are inside the fluid and their nature can
be obtained by a further study.
 \vspace{0.8cm}

\scalebox{0.25}{\includegraphics{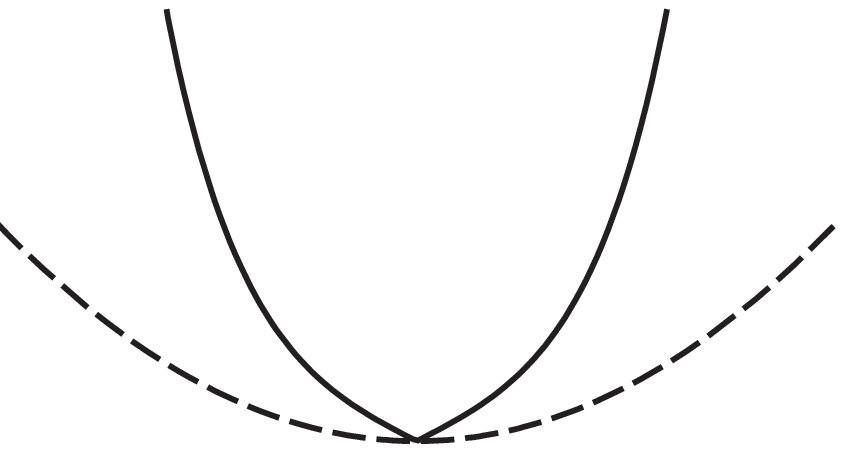}} \hspace{0.3cm}
\scalebox{0.25}{\includegraphics{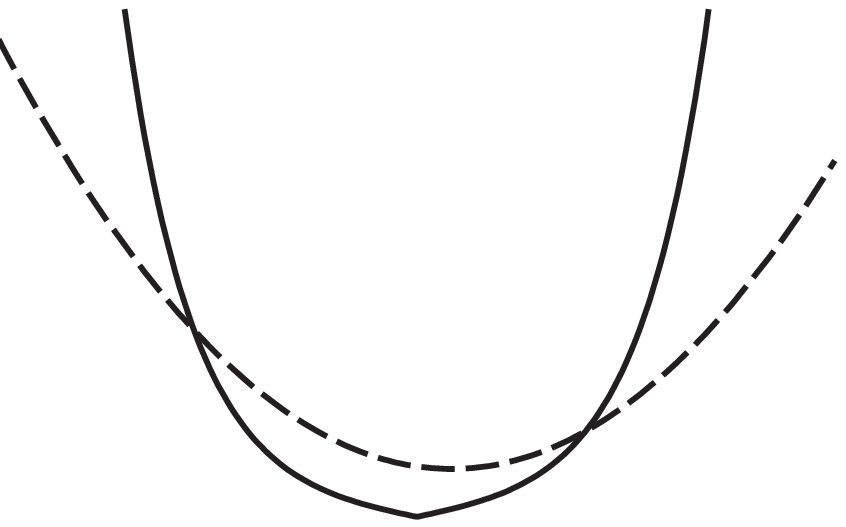}}\hspace{0.3cm}
\scalebox{0.25}{\includegraphics{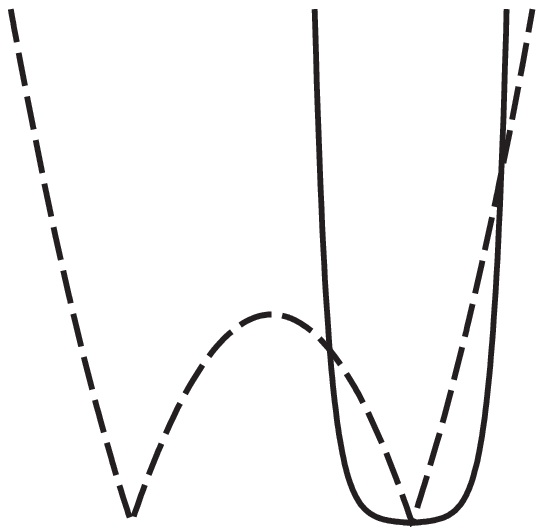}}\hspace{0.3cm}
\scalebox{0.25}{\includegraphics{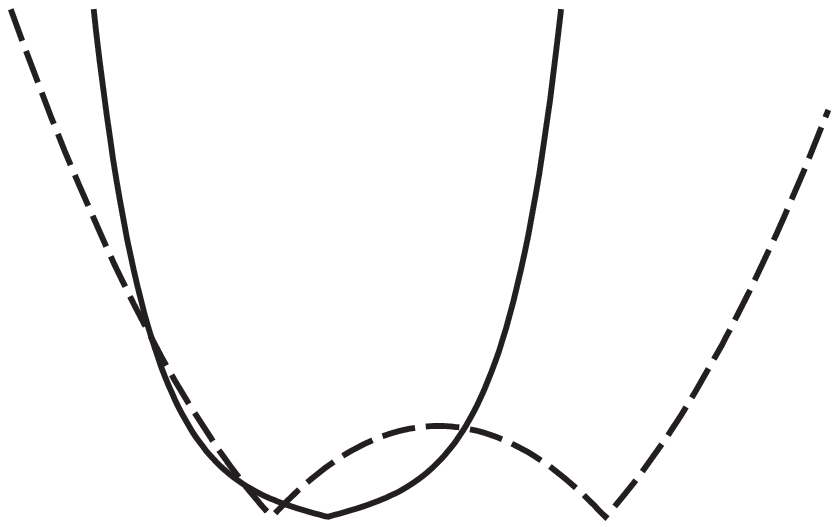}}\hspace{0.3cm}
\scalebox{0.25}{\includegraphics{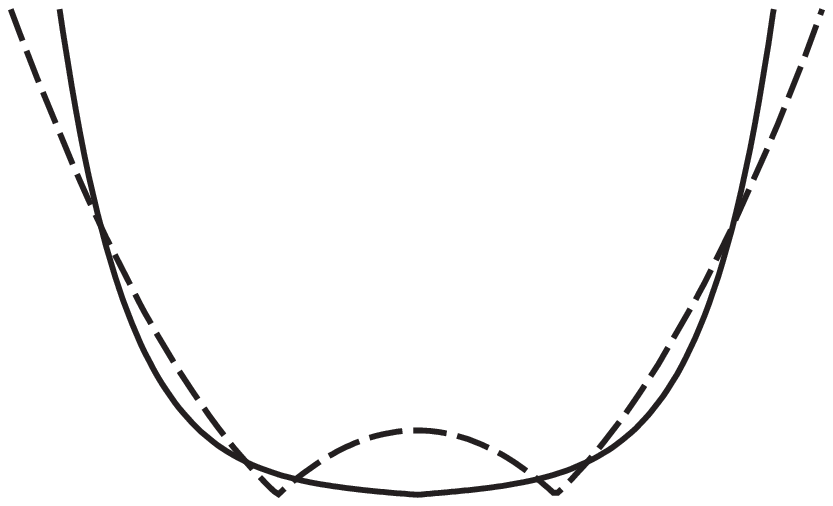}}
\\

\hspace{2cm} \scalebox{0.40}{\includegraphics{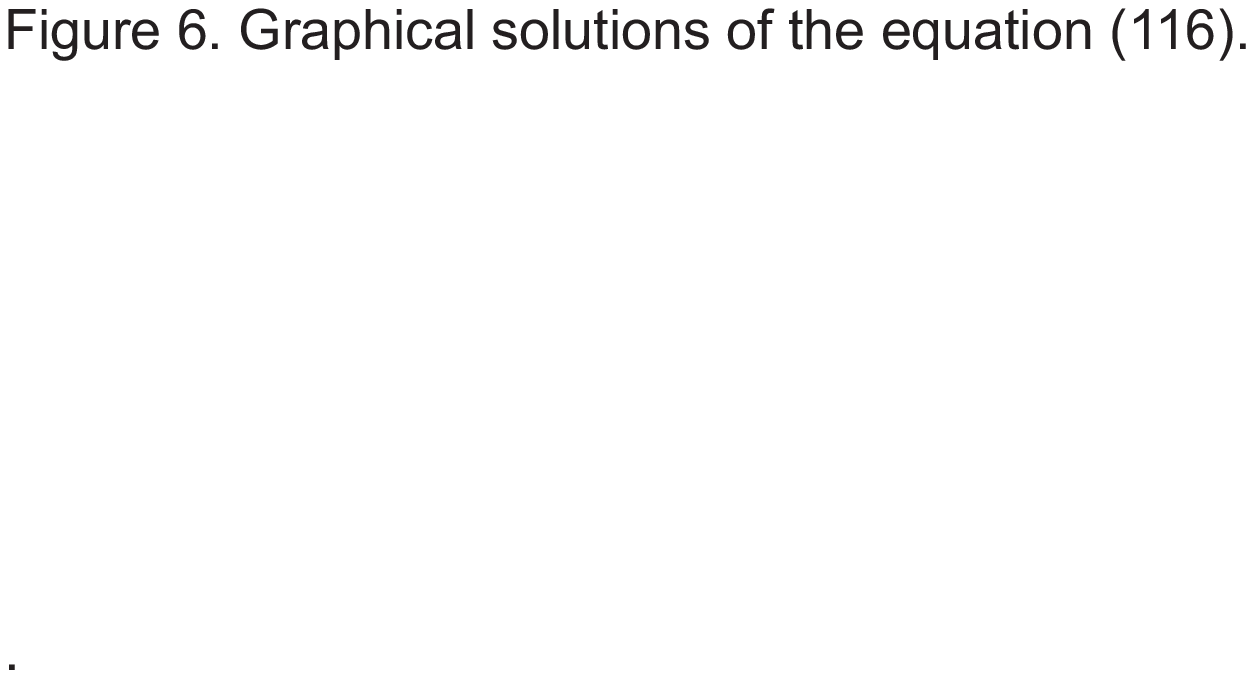}}

\end{document}